\documentclass[preprint,showpacs,preprintnumbers,amsmath,amssymb]{revtex4}

\usepackage{graphicx}
\usepackage{dcolumn}
\usepackage{bm}
\usepackage{epsfig}

\begin{document}

\preprint{APS/123-QED}

\title{A Particle In Cell code development for high current ion beam transport and plasma simulations }

\author{N. Joshi}
\email{ni.joshi@gsi.de}

\affiliation{
Goethe University, Frankfurt,   Germany \\
GSI Helmholtzzentrum for Scherionenforschung GmbH, Germany \\
}%

\date{August 20, 2015}

\begin{abstract}

A simulation package employing a Particle in Cell (PIC) method is developed to study the high current beam transport and the dynamics of  plasmas.
This package includes subroutines those are suited for various planned projects at  University of Frankfurt.
In the framework of the storage ring project (F8SR) the code was written to describe the beam optics in  toroidal magnetic fields.
It is used to design an injection system for a ring with closed magnetic field lines.
The generalized numerical model, in Cartesian coordinates is used to describe the intense ion beam transport through the chopper system in the low energy beam section of the FRANZ project.
Especially for the chopper system, the Poisson equation is implemented with irregular geometries.
The Particle In Cell model is further upgraded with a Monte Carlo Collision subroutine for simulation of plasma in the volume type ion source.

\end{abstract}

\pacs{29.20.db,29.27.Eg,41.75.-i}
\maketitle

\section{\label{sec:level1}Introduction\protect\\
 }

An increasing number of projects are dealing with high intensity accelerators and plasma devices.
An efficient simulation tool for charged particle transport in such machines is in demand.
Institut f\"ur Angewandte Physik at University Frankfurt is engaged in many such projects.

The storage ring project known as ``Figure-8 Storage Ring'' (F8SR) deals with accumulation of low energy high current beams \cite{F8SR}. 
Particularly, a proton beam at an energy of $150~keV$ with a current of about $200~mA$ is desired to inject into the ring comprising of toroidal like magnetic field with closed magnetic field lines.
The preliminary experiments are designed to operate at room temperature with an aim to build an injection system.
A numerical model was required to calculate the beam optics taking into account effect of fringing fields without paraxial approximation and momentum conservation in magnetic field configuration.
The particle trajectories must be calculated in complex magnetic field.
The particle motion is dominated by various kinds of drifts arising due to inhomogeneity in the field distribution.
The PIC model was implemented in circular toroidal coordinate systems.

The FRAnkfurt Neutron source at Stern-Gerlach Zentrum (FRANZ) is an ambitious project that aims to produce a high density neutron flux \cite{FRANZ}.
The neutrons will be produced using $^7Li(p,n)^7Be$ reaction, by accelerating intense proton beam (about $200~mA$) up to an energy of $2.1~MeV$ on the $^7Li$-target.
The beam transport becomes challenging especially in the view of high perveance proton beam.
In the low energy beam transport (LEBT) section a beam chopper is desired to avoid beam loading in the RFQ.
The simulation tool was designed to calculate space charge dominated beam transport in the chopper system taking into account effects of secondary electron production, dynamical effects on the bunched beam, and space charge compensation.
The Poisson equation was modified with irregular boundary condition to simulate the electric field from curved plates.
The effect of mirror charges and sparking problem can be investigated. 

The benchmarking of the simulation code is an important factor in gaining the confidence.
The model validation was done by comparison of simulation results with measurements.
The PIC model in toroidal coordinate was verified by beam transport experiments in toroidal magnetic fields.
The simulation results showed a good agreement within an acceptable error range.
The initial results from the kicker experiments are compared with the simulation result those shows the effect of the secondary electrons on beam diagnostics.

A subroutine based on Monte Carlo technique is incorporated to study the production mechanism of different ion species in simple He- plasma generated in the volume type ion source.
In the following section we will discuss the development of the numerical model and the special features implemented in the package for various scenario.

\section{Particle in Cell method}
\label{Particle in Cell method}
The main task in charge particle simulation is to calculate interparticle Coulombic forces.
An intuitive, simplest model is so called Particle Particle (PP) method which calculates the electric force on each particle due to others in every time step.
This has limits not only in terms of execution time but also the data management with respect to number of particles under consideration.
To overcome these limitations Particle in Cell method can be used.
In PIC method the information of particle distribution is transferred to the set of grid points.

\begin{figure}[!h]
\vspace{0.5cm}
\begin{center}
\includegraphics*[width=80mm]{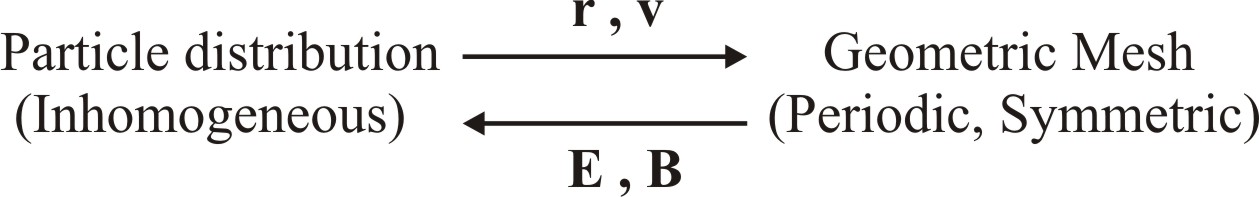}
\end{center}
\caption{PIC methodology.}
\label{PIC_philosophy}
\end{figure}

\begin{figure}[!h]
\vspace{0.5cm}
\begin{center}
\includegraphics*[width=8cm]{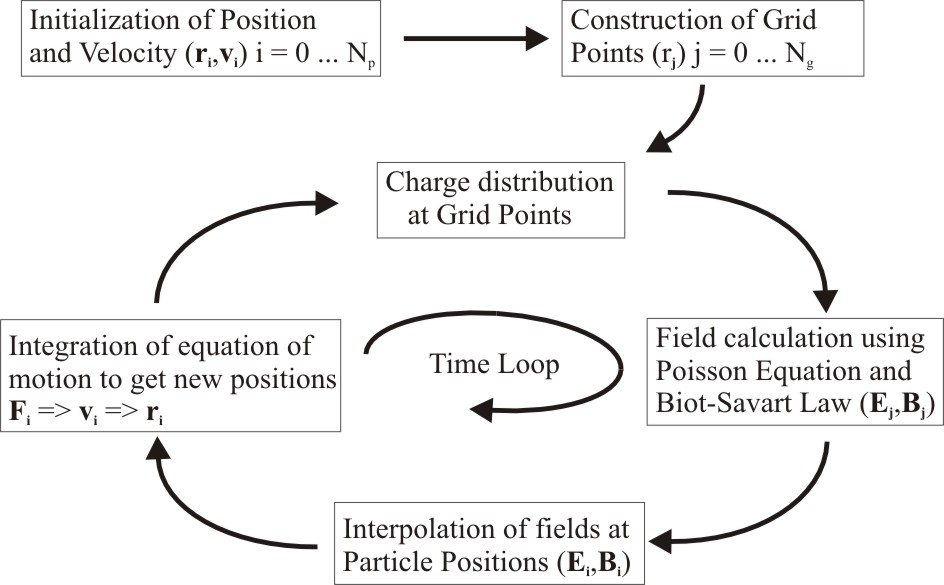}
\end{center}
\caption{Flow chart of the PIC algorithm.}
\label{PIC_algorithm}
\end{figure}

The grid can be conveniently chosen according to the system geometry e.g. cartesian coordinates in the simplest case, or a cylindrical mesh for circular symmetry.
The electric field is calculated on grid points using Poisson equation.
The periodicity and the symmetry of the grid give decisive advantage in solving the Poisson equation.
This electric field is interpolated back on the particle positions \cite{BirdsallLangdon}.
The general algorithm for PIC method is shown in Fig.~\ref{PIC_algorithm}.

\subsection{Initial Particle distribution}
\label{Initial Particle distribution}

Different approaches are implemented to define the initial phase space distribution of particles.
Commonly in the accelerator physics Kapchinsky-Vladimirsky (KV), Waterbag or Gaussian distribution are used to define the beam like distribution with main component of momentum in the forward direction with a small spread in transverse velocity plane.
On the other hand the Maxwellian distribution is defined for a thermalized distribution.
It is convenient also to use a homogenous distribution for analytical comparison with a numerical model e.g. simulated potential distribution due to a homogenous cylindrical beam which can easily be compared using Gauss' Law.

\subsection{Charge division}
\label{Charge division}

For an efficient calculation of inter particle forces the PIC model was used.
Cartesian, cylindrical, or toroidal grids, as per requirement imposed by geometry were generated.
The first order weighing scheme was used to calculate the charge density at grid points.
Each of the particles is identified in a particular cell and then the charge is attributed to grid points according to a relative volume in the 3D space.

\begin{figure}[!h]
\vspace{0.5cm}
\begin{center}
\includegraphics*[width=60mm]{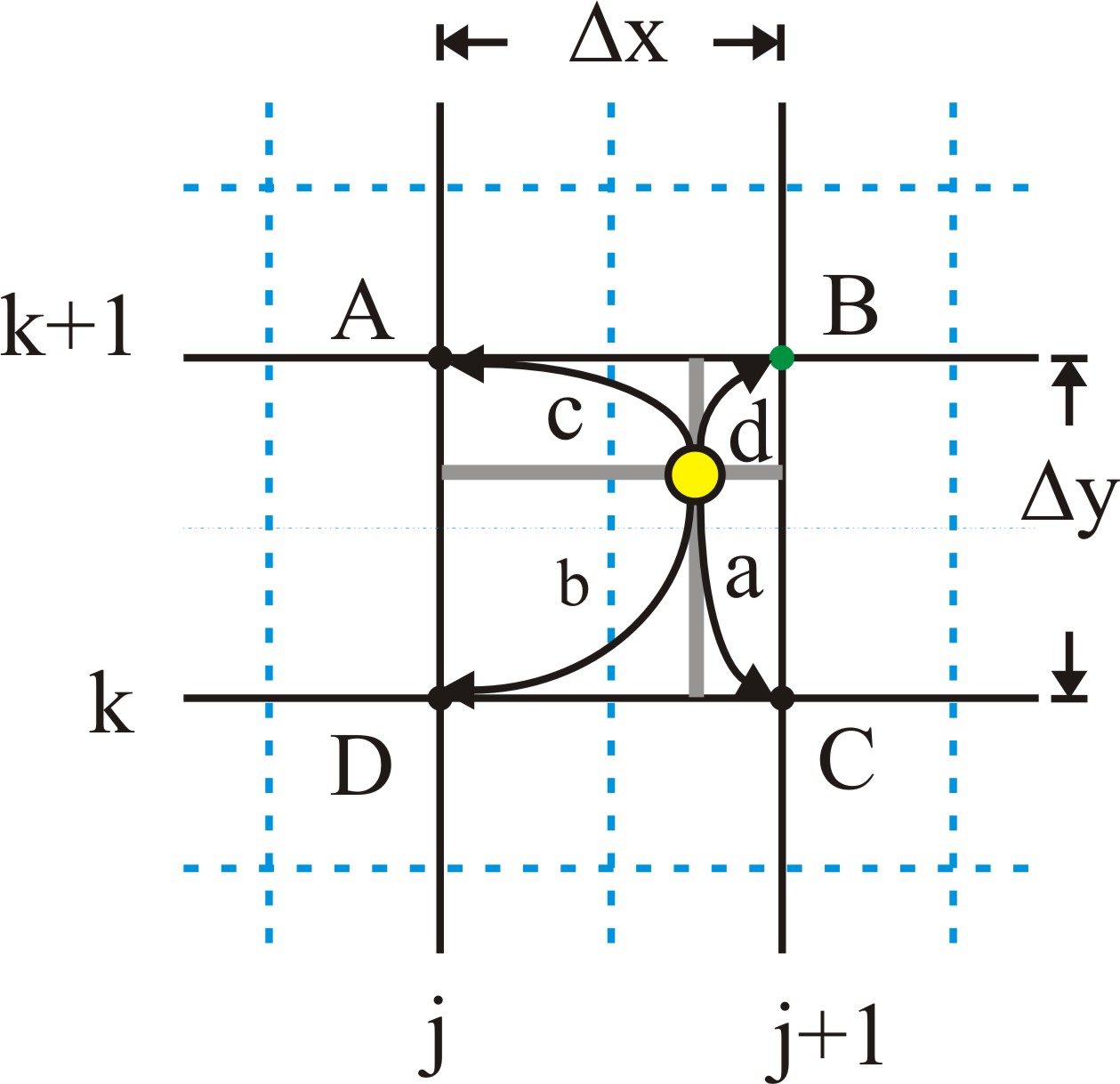}
\end{center}
\caption{PIC charge distribution in cartesian.}
\label{PICcdcart}
\end{figure}

For example, as shown in Fig.~\ref{PICcdcart}, in 2-dimensional case, a particle is identified at grid point B called a Nearest Grid Point (NGP).
The charge of this particle, which can be a macro particle with cluster of particles, is divided according to the inverse area weight.
Charge density at the grid point is given by,

\begin{equation}
Q_B=Q_0 \frac{area~(b)}{area(ABCD)},
\end{equation}

and so on for $Q_A,Q_C,Q_D$ where $Q_0$ is the macro charge of a single particle also called \textit{super particle}.
Thus the nearest grid point, point B, is weighted maximum as compared to the point D.

\subsection{Poisson equation}
\label{Poisson equation}

Poisson equation,
	
	\begin{equation}
	\nabla^{2}\phi(\mathbf r)=-\frac{\rho(\mathbf r)}{\epsilon_{0}}
	\label{PoissonEq}
	\end{equation}

is written in cartesian coordinates as,

	\begin{equation}
	\frac{\partial^{2} \phi}{\partial x^{2}} +\frac{\partial^{2} \phi}{\partial y^{2}}+\frac{\partial^{2} \phi}{\partial z^{2}}=-\frac{\rho}{\epsilon_{0}}
	\label{PoissonEq_cart}.
	\end{equation}

This equation is discretized as

	\begin{eqnarray}
	\frac{\phi_{i+1,j,k}-2\phi_{i,j,k}+\phi_{i-1,j,k}}{\Delta x^2}+\frac{\phi_{i,j+1,k}-2\phi_{i,j,k}+\phi_{i,j-1,k}}{\Delta y^2}\nonumber\\
	+\frac{\phi_{i,j,k+1}-2\phi_{i,j,k} 
	+\phi_{i,j,k-1}}{\Delta z^2}=-\frac{\rho_{i,j,k}}{\epsilon_{0}}. \nonumber\\
	&
	\label{PoissonEq_discrete}
	\end{eqnarray}

\begin{figure}[!h]
\vspace{0.5cm}
\begin{center}
\includegraphics*[width=8cm]{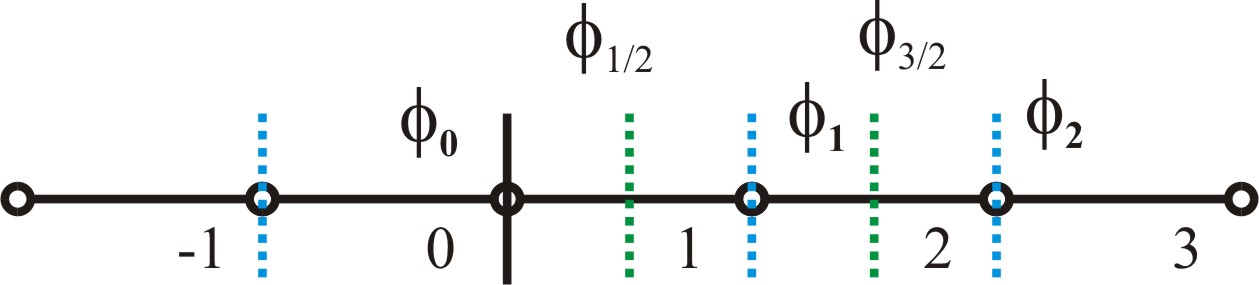}
\end{center}
\caption{Numerical stencil for 1-D equi-distant grid describing discretization of Poisson equation.}
\label{stencil_1}
\end{figure}

In this scheme the gradient is taken at half step as shown in Fig.~\ref{stencil_1}.

Simplifying Eq. (\ref{PoissonEq_discrete}) we get,

	\begin{eqnarray}
\frac{1}{\Delta z^2} \phi_{i,j,k+1} + \frac{1}{\Delta y^2} \phi_{i,j+1,k} + \frac{1}{\Delta x^2} \phi_{i+1,j,k} \nonumber\\
- \left(\frac{2}{\Delta x^2}+\frac{2}{\Delta y^2}+\frac{2}{\Delta z^2} \right)  \phi_{i,j,k} +             \frac{1}{\Delta x^2} \phi_{i-1,j,k} \nonumber\\
+ \frac{1}{\Delta y^2} \phi_{i,j-1,k} + \frac{1}{\Delta z^2} \phi_{i,j,k-1} =-\frac{\rho_{i,j,k}}{\epsilon_{0}}
	\label{PoissonEq_matrix}.
	\end{eqnarray}

This gives a matrix equation 

	\begin{equation}
	\mathbf{A}\cdot {\phi_{i,j,k}} =\frac{{\rho_{i.j.k}}} {\epsilon_{0}},
	\label{Eq_matrix_phi}
	\end{equation}

of the form 

	\begin{equation}
	\mathbf{A}\cdot \mathbf{x} =\mathbf{b}
	\label{Eq_matrix_A}.
	\end{equation}

The matrix $A$ consist of a numerous of zero elements, and is  known as \textit{sparse}.
The sparse matrix format allows efficient use of computational memory by storing only non zero values and the corresponding reference pointer.
After having the matrix equations with above described form, the iterative methods are used to solve them.
They are effective when the number of equations $N>10^6$, where $N$ is the number of grid points.
The Biconjugate gradient stabilized (BiCGSTAB) method is used to solve a non-symmetric linear system.

\subsection{Boundary conditions and multigrid method}
\label{Boundary conditions and multigrid method}

The matrix equation requires initial values also called boundary conditions.
In the case of open boundary condition the potential at the selected grid points, normally on the surface, is calculated using Coulomb interaction, setting initial conditions for the system.

\begin{figure}[!h]
\vspace{0.5cm}
\begin{center}
\includegraphics*[width=7cm]{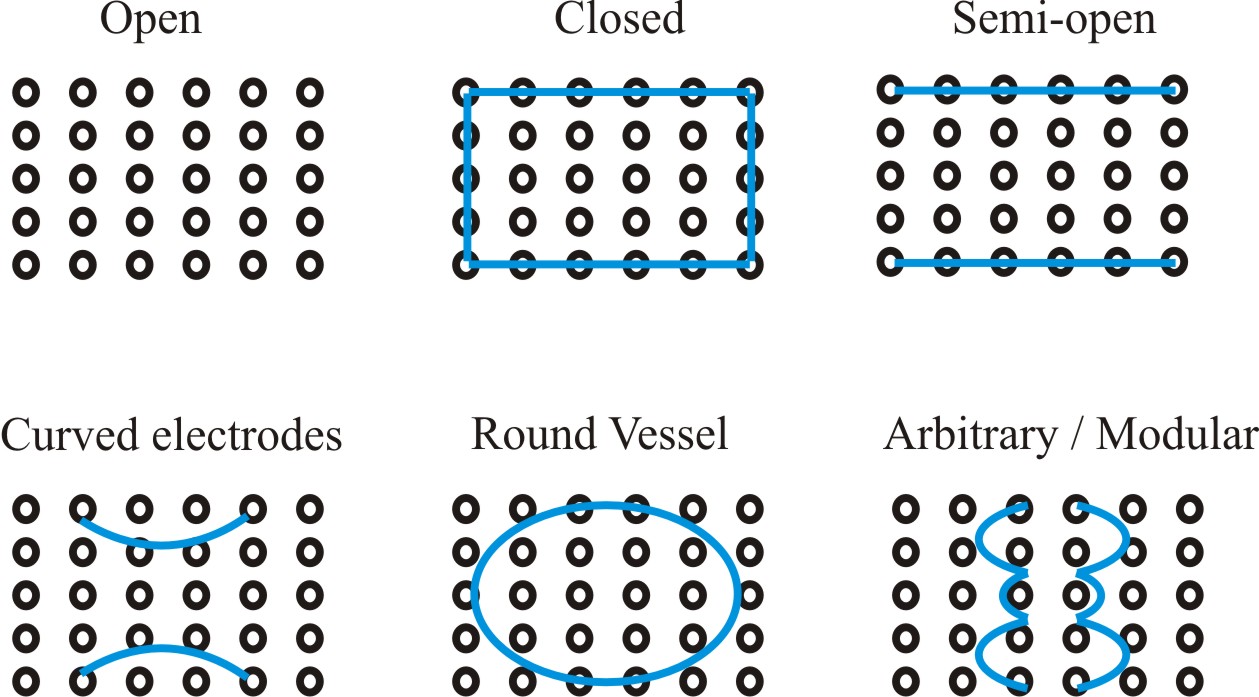}
\end{center}
\caption{Different type of boundary conditions implemented in the model.}
\label{boundary}
\end{figure}

The closed boundary conditions are more natural, with boundaries defined with respect to system geometry e.g the vessel wall.
The semi-open or semi-closed boundary condition is the combination of the both, e.g. the vessel defined in transverse direction and open boundary in longitudinal.
The code was further evolved to include so called arbitrary boundary condition. 
This facilitates definition of curved vessel boundary in Cartesian coordinates or vice versa (see Fig.~\ref{boundary}).

\subsubsection{Poisson Equation with arbitrarily defined boundary condition}
\label{Poisson Equation with arbitrarily defined boundary condition}

\begin{figure}[!h]
\vspace{0.5cm}
\begin{center}
\includegraphics*[width=8cm]{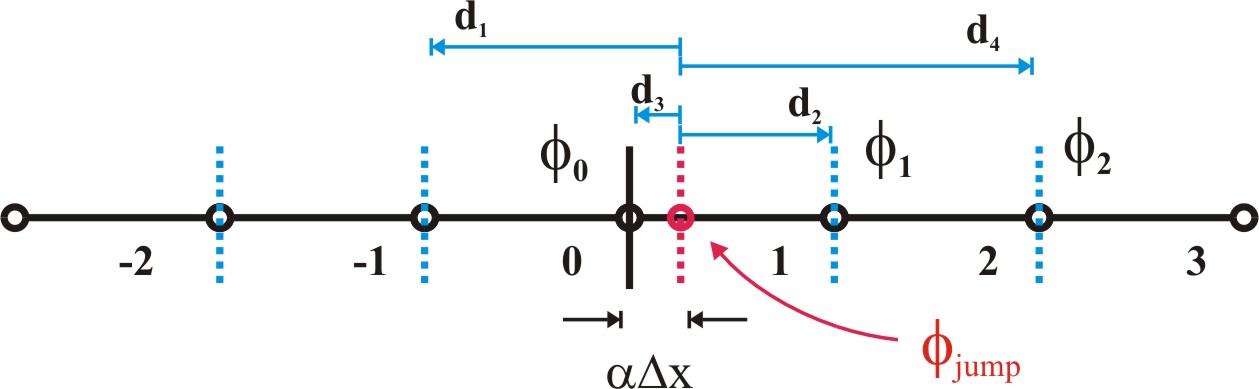}
\end{center}
\caption{Numerical stencil for 1-D equi-distant grid with a potential defined at an arbitrary distance from a grid point.}
\label{stencil_2}
\end{figure}

In case of the irregular geometry where the fixed potential value is not defined exactly on the mesh point the matrix elements need correction \cite{Johansen_Colella} \cite{Jomaa}.
The numerical stencil in 1-D case is shown in the Fig.~\ref{stencil_2}.

The gradient at $\phi_{1/2}$ is then written by fitting quadratic polynomial as

	\begin{equation}
	\phi'_{1/2}=\frac{1}{\Delta x} \left[ \left(\phi_{1}-\phi_{jump} \right) \frac{d_1}{d_2}- \left(\phi_{2}-\phi_{jump} \right) \frac{d_3}{d_4} \right]
	\label{phi_jump}.
	\end{equation}

Putting the values of $d$'s

	\begin{eqnarray}
	d_1 = \Delta x+ \alpha \Delta x, \hspace{1.0cm}  d_2=\Delta x- \alpha \Delta x, \nonumber \\
	d_3  =  \alpha \Delta x, \hspace{1.0cm}  d_4=2\Delta - \alpha \Delta x,
	\label{ds}
	\end{eqnarray}

in Eq. (\ref{phi_jump}) we get,

\small
	\begin{eqnarray}
\phi'_{1/2}= \frac{1}{\Delta x} \left[ -\frac{2}{(1-\alpha)(2-\alpha)} \phi_{jump} + \left(\frac{1+\alpha}{1-\alpha} \right) \phi_{1}- \left(\frac{\alpha}{2-\alpha}\right) \phi_{2} \right].& \nonumber\\
&
	\label{phi_dash}
	\end{eqnarray}
\normalsize

Putting expression from equation (\ref{phi_dash}) for $\phi'_{1/2}$ in standard second derivative

	\begin{equation}
	 \phi''_{1}= \frac {1}{\Delta x} \left(\phi'_{3/2}- \phi'_{1/2} \right)
	\label{phi_ddash}.
	\end{equation}
\small
	\begin{eqnarray}
 	\phi''_{1}= \frac {1}{\Delta x^2} \left[ \frac{2}{(1-\alpha)(2-\alpha)}  \phi_{jump}- \left( \frac{2}{1-\alpha} \right) \phi_{1}+\left(\frac{2}{2-\alpha} \right) \phi_{2} \right]. \nonumber \\
	&
	\label{phi_ddash_l}
	\end{eqnarray}
\normalsize

Above described scheme is useful for a boundary condition, where the $\phi_{jump}$ is defined left to the $\phi_1$. 
In the case where the $\phi_{jump}$ is defined right to the $\phi_1$, a similar analysis can be carried out to give,

\small
	\begin{eqnarray}
	 \phi''_{3/2}= \frac {1}{\Delta x^2} \left[ \frac{2}{(1-\alpha)(2-\alpha)}  \phi_{jump}- \left( \frac{2}{1-\alpha} \right) \phi_{1}+\left(\frac{2}{2-\alpha} \right) \phi_{0} \right] . \nonumber\\
	&
	\label{phi_ddash_r}
	\end{eqnarray}
\normalsize

This numerical technique can be extended in 2D or 3D easily.
Then it can be used to calculate the potential arising from the curved boundary in the cartesian coordinates.

For the left boundary then the Poisson equation can be written as,

\small
	\begin{eqnarray}
\frac {2}{(2-\alpha_{z})~\Delta z^2} \phi_{i,j,k+1} + \frac {2}{(2-\alpha_{y})~\Delta y^2} \phi_{i,j+1,k} + \frac {1}{\Delta x^2} \phi_{i+1,j,k}  \nonumber \\
-(\frac {2}{\Delta x^2}+\frac {2}{(1-\alpha_{y})~\Delta y^2}+\frac {2}{(1-\alpha_{z})~\Delta z^2}) \phi_{i,j,k}+ \frac {1}{\Delta x^2} \phi_{i-1,j,k} \nonumber \\
=-\frac{\rho_{i,j,k}}{\epsilon_0} - 2 \phi_{jump} \left[ \frac{1}{(1-\alpha_z)(2-\alpha_z) ~\Delta z^2} + \frac{1}{(1-\alpha_y)(2-\alpha_y) ~\Delta y^2} \right],\nonumber \\
&
	\end{eqnarray}
\normalsize
and for the right boundary condition we have,

\small
	\begin{eqnarray}
\frac {2}{(2-\alpha_{z})~\Delta z^2} \phi_{i,j,k-1} + \frac {2}{(2-\alpha_{y})~\Delta y^2} \phi_{i,j-1,k} + \frac {1}{\Delta x^2} \phi_{i+1,j,k}  \nonumber \\
-(\frac {2}{\Delta x^2}+\frac {2}{(1-\alpha_{y})~\Delta y^2}+\frac {2}{(1-\alpha_{z})~\Delta z^2}) \phi_{i,j,k}+ \frac {1}{\Delta x^2} \phi_{i-1,j,k} \nonumber \\
=-\frac{\rho_{i,j,k}}{\epsilon_0} - 2 \phi_{jump} \left[ \frac{1}{(1-\alpha_z)(2-\alpha_z) ~\Delta z^2} + \frac{1}{(1-\alpha_y)(2-\alpha_y) ~\Delta y^2} \right] .\nonumber \\
&
	\end{eqnarray}
\normalsize

Please note here, this model was used for Wien type chopper system. Hence only the cases of jump in $y-$ and $z-$ directions are analyzed (refer section \ref{Chopper system for FRANZ}).
The electric field on mesh points then can calculated simply by,

	\begin{equation}
	\mathbf{E}= -\nabla \phi.
	\end{equation}

Each component of the electric field can then be written in descretized form as,

	\begin{eqnarray}
	 E_x(i,j,k) &= & - \frac{\phi_{i+1,j,k}-\phi_{i-1,j,k}}{2 \Delta x}  , \nonumber \\
	 E_y(i,j,k) &= & -\frac{\phi_{i,j+1,k}-\phi_{i,j-1,k}}{2 \Delta y}   , \nonumber \\
	 E_z(i,j,k) &=&  -\frac{\phi_{i,j,k+1}-\phi_{i,j,k-1}}{2 \Delta z} . \nonumber \\  
&
	\end{eqnarray}

Consequently, these electric fields at grid points are interpolated back on particle positions employing reversed PIC scheme \cite{HockneyEastwood}.

\subsubsection{Time evolution}
\label{Time evolution}

The evolution of the particle is then calculated by simply putting fields and positions in non relativistic Lorentz equation,
	\begin{equation}
	F= \frac{dp}{dt} =q (\mathbf{E}+[\mathbf{v} \times \mathbf{B}]).
	\end{equation}
The time step $dt$ plays an important role.
For example, to describe the charged particle motion in magnetic field $dt$ must be chosen at least $20$ times lesser than the gyration period.
To describe the motion of particle over a longer time guiding centre equations must be used and symplectic integrators have to be involved.

\section{Transport in Toroidal Magnetic Fields}
\label{Transport in Toroidal Magnetic Fields}

Storage ring with Figure-8 geometry will be built for accumulation of intense low energy ion beams.
The structure uses continuous longitudinal magnetic field (up to $ 6-8~T$) for focusing the beams in the ring.
The room temperature experiments are aimed to build an injection system with two toroidal segments with maximum on axis field of $0.6~T$ \cite{EPAC_2008} .
The dynamics of charged particle beams in magnetic fields is characterized by a gyrating motion.
In curved magnetic fields the ion beam is guided on a circular path, additionally dominated by drift motion, namely $\mathbf{R} \times \mathbf{B}$ drift due to curved magnetic field, $\nabla\mathbf{B} $ drift due to the inhomogeneous field and $\mathbf{E} \times \mathbf{B}$ drift due to the space charge \cite{Chen}.
The PIC code was designed to describe the beam transport in curved sectors.
Main features of the PIC model implemented in this case are

\begin{itemize}
	\item Poisson equation in circular toroidal coordinates
	\item Complete 3D space charge routine, space-charge compensation through compensation electrons (CE) trapped in magnetic field
	\item Simulation results can directly be compared with experimentally available optical detection system
	\item Inclusion of secondary electrons (SE) produced on wall due to beam losses
\end{itemize}

\begin{figure}[!h]
\vspace{0.5cm}
\begin{center}
\includegraphics*[width=80mm]{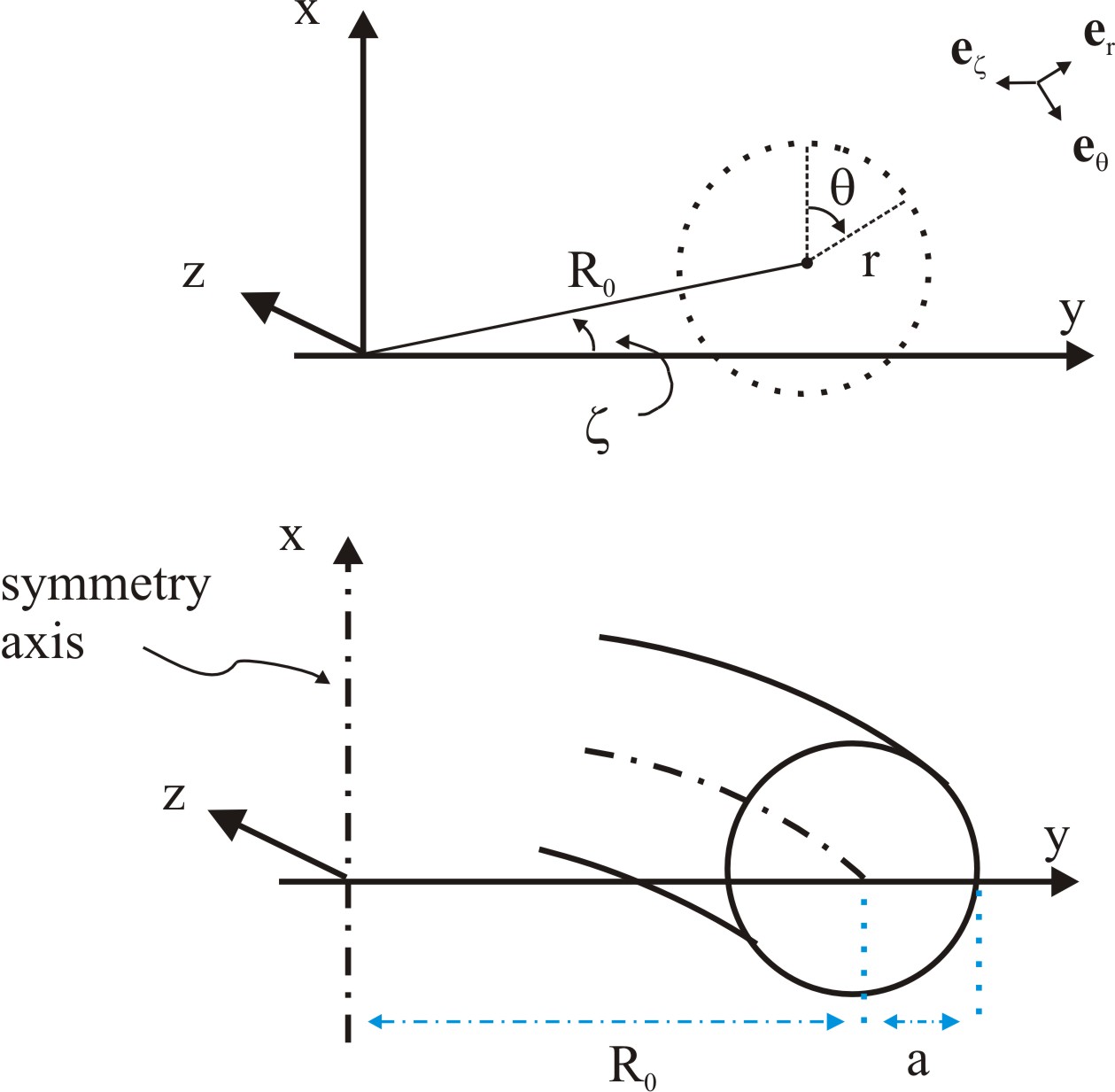}
\end{center}
\caption{Circularly symmetric toroidal coordinates.}
\label{stellar_1}
\end{figure}

For the model the standard toroidal coordinates ($r, ~\theta,~ \zeta$) were used and the Poisson equation was discretized in the same (see Fig.~\ref{stellar_1}) \cite{Balescu}.

	\begin{itemize}
	\item $r$ : minor radius of toroidal segment;
	\item $\theta$ : the poloidal angle measured in $x-y$ plane from $+ve~x$-axis;
	\item $\zeta$ : the toroidal angle measured in $y-z$ plane from $+ve~y$-axis.
	\end{itemize}

Special care has to be taken with respect to boundary conditions.
In $\zeta$ direction open boundaries were defined and in $\theta$ direction periodic boundary.
In radial direction the potential is set to zero at $r=a$ where $a$ is vessel radius.
For the second boundary condition at $r=0$ , the Gauss's law is used.
	\begin{equation}
	\int E\cdot dS =  \frac{Q_{enclosed}}{\epsilon_{0}}.
	\label{gauss}
	\end{equation}

\subsection{Experiments}
\label{Experiments_MSR}

\begin{figure}[!h]
\vspace{0.5cm}
\begin{center}
\includegraphics*[width=8cm]{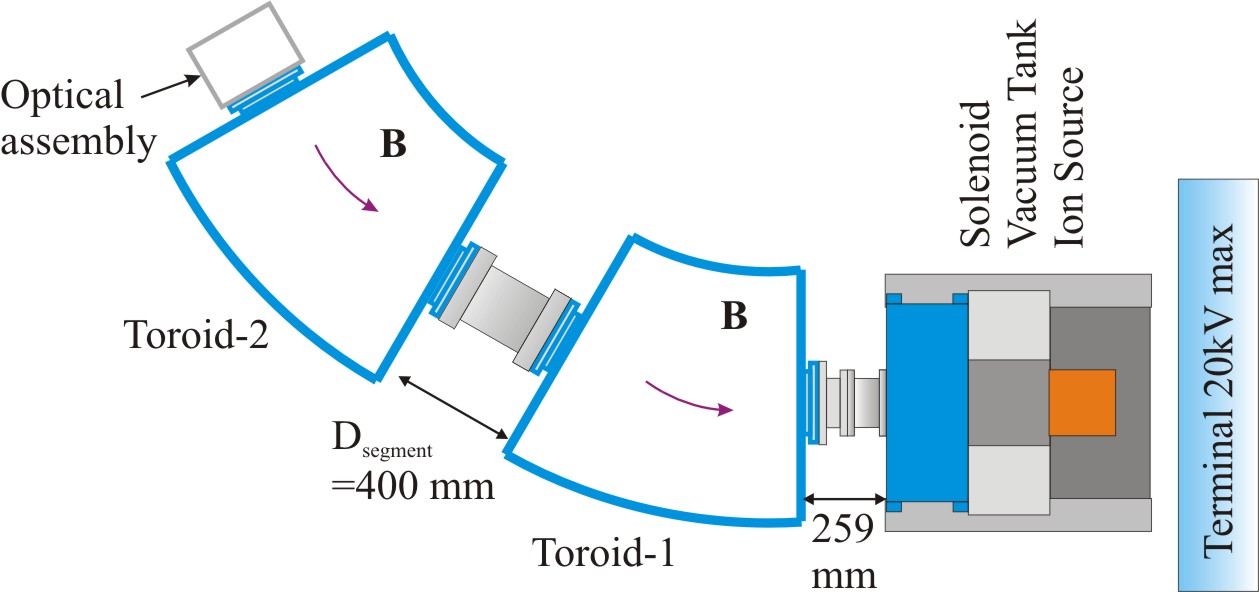}
\end{center}
\caption{Schematic diagram of the top view of the experimental setup}
\label{stellar_2}
\end{figure}

Fig. \ref{stellar_2} shows the current experimental setup comprising two toroidal segments.
This forms the main beam line for injection experiments.

\begin{figure}[!h]
\vspace{0.5cm}
\begin{center}
\includegraphics*[height=40mm,width=80mm]{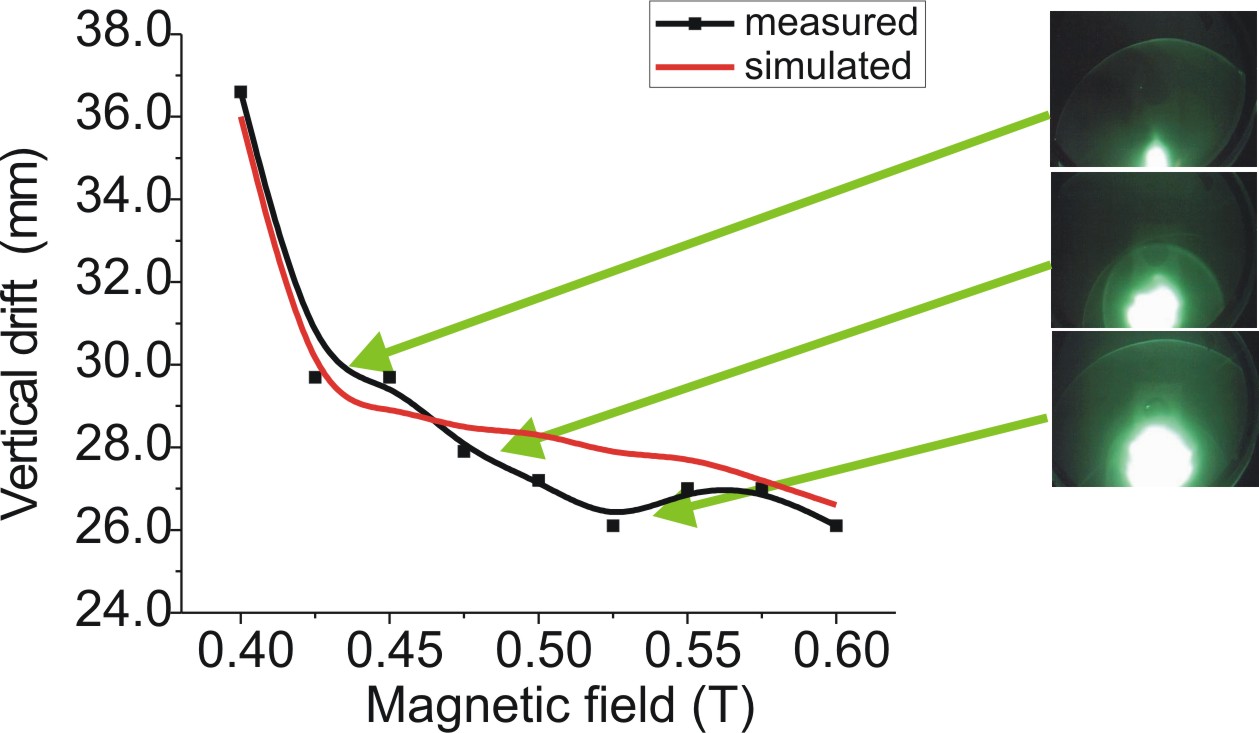}
\end{center}
\caption{Measured vertical drift ($ \mathbf{R} \times \mathbf{B}$) as a function of magnetic field compared with simulation. The example images show proton beam at $12~keV$ in three different magnetic field strength of $0.4~T$,$0.45~T$ and $0.50~T$. }
\label{stellar_3}
\end{figure}

In the early stages beam transport experiment were performed with fixed diagnostics probe at the end of first segment.
Fig.~\ref{stellar_3} shows the proton beam with energy $12~keV$ detected using a phosphor screen.
The beam centre is seen to be shifted vertically upward nreare to axis, as a function of the increasing magnetic field.
The simulation results and measurements are compared in the graph.

\begin{figure}[!h]
\vspace{0.5cm}
\begin{center}
\includegraphics*[height=40mm,width=80mm]{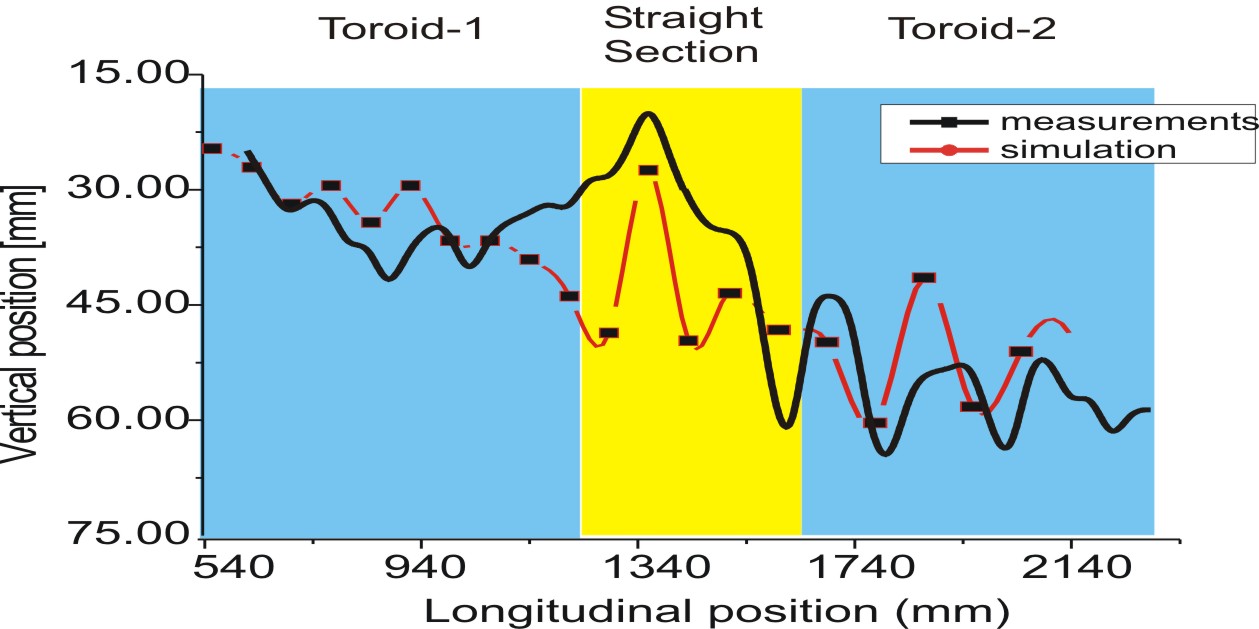}
\end{center}
\caption{The vertically drifted beam position plotted as a function of longitudinal axis.}
\label{stellar_4}
\end{figure}

In the current experimental setup two segments are coupled to form a ripple like structure. 
Fig.~\ref{stellar_4} shows the vertically drifted beam position along the beam path in longitudinal direction.
The simulation results are qualitatively in good agreement with measurement.
The error largely occurs in the straight section due to heavy beam losses producing background noise of secondary electrons.
Details of the experiments are discussed in \cite{IPAC_2010}.

\section{Chopper system for FRANZ}
\label{Chopper system for FRANZ}

\begin{figure}[!h]
\vspace{0.5cm}
\begin{center}
\includegraphics*[width=80mm]{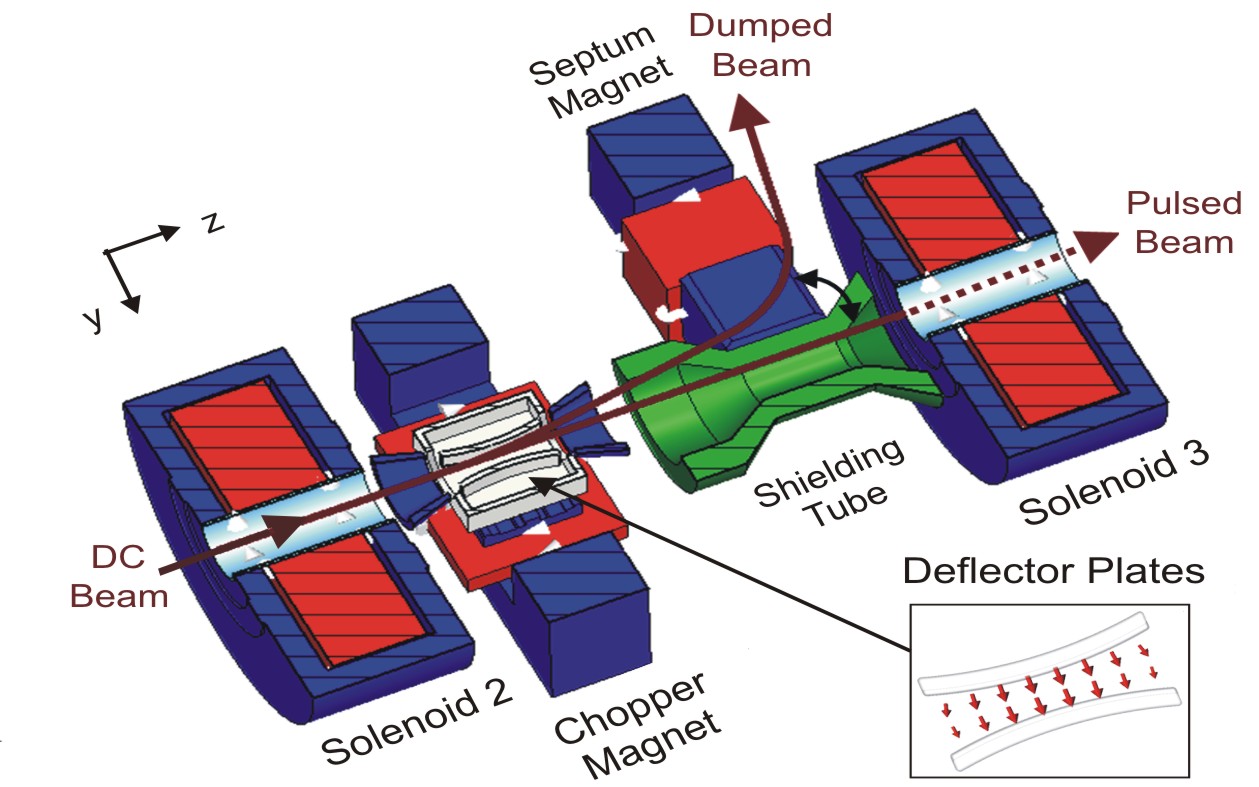}
\end{center}
\caption{Concept of Wien type chopper system.}
\label{wien}
\end{figure}

A Wien-type kicker system with crossed electric-magnetic fields ($\mathbf{E} \times\mathbf{B}$) was proposed \cite{Wiesner}\cite{PAC_2009} for FRANZ project in low energy beam transport section.
A static magnetic field deflects the beam into the beam dump.
A pulsed electric field produced using curved plates compensates the deflection for a short time period and injected into the RFQ (see Fig.~\ref{wien}).

Transport of high current beam at low energy is dominated by space charge.
The beam potential at a focus can be as high as $1600~V$.
To reduce beam losses due to the space charge forces the chopper system has to be compact.
The PIC code is implemented for beam tracking, to investigate beam power dissipation and effects of secondary electrons on the beam transport.

Main features of the PIC model in this case are

\begin{itemize}
	\item Semi open boundary condition that facilitates definition of circular flanges at the entrance and variable aperture (Iris) at the output
	\item Continual generation and destruction of ions to mimic input dc beam and the beam dump
	\item Use of modular boundary condition to simulate curved deflector plates and circular repeller ring electrode in the Cartesian coordinates
\end{itemize} 

This simulation tool describes the evolution of beam trajectory over the time.
Multiple species can be transported simultaneously, thus effects of electrons on proton beam can be investigated.
\begin{figure}[!h]
\vspace{0.5cm}
\begin{center}
\includegraphics*[width=80mm]{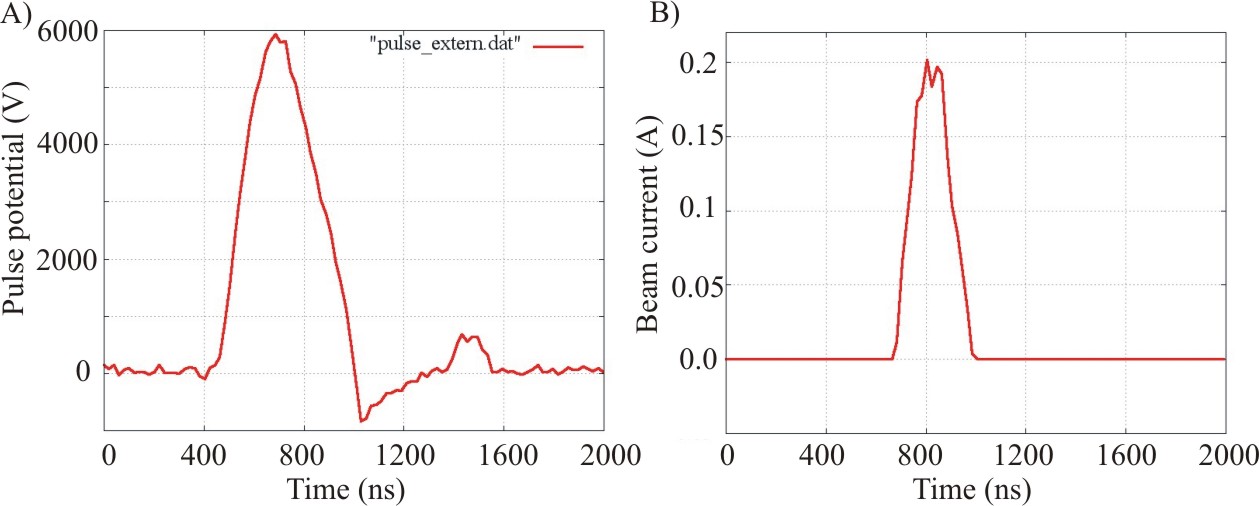}
\end{center} 
\caption{ A) Graph showing the measured input pulse B) Simulated beam current at the chopper exit.}
\label{pulse_current}
\end{figure}
Fig.~\ref{pulse_current} shows the signal from pulse generator and the simulated proton beam current downstream of the chopper system. 
A proton beam with an energy $120~keV$ and beam current $200~mA$ was transported through an aperture of $40~mm$ over the distance of $780~mm$.
No space charge compensation was assumed.
To show the effects of space charge compensation through secondary electrons a smaller aperture of $35~mm$ was defined.
The electrons were defined to produce at the beam dump and on the front wall.
As the pulse rises, some electrons are sweeped with the ion beam into the aperture due to the beam potential.
This causes about $20\%$ of the space charge compensation.
\begin{figure}[!h]
\vspace{0.5cm}
\begin{center}
\includegraphics*[height=30mm,width=60mm]{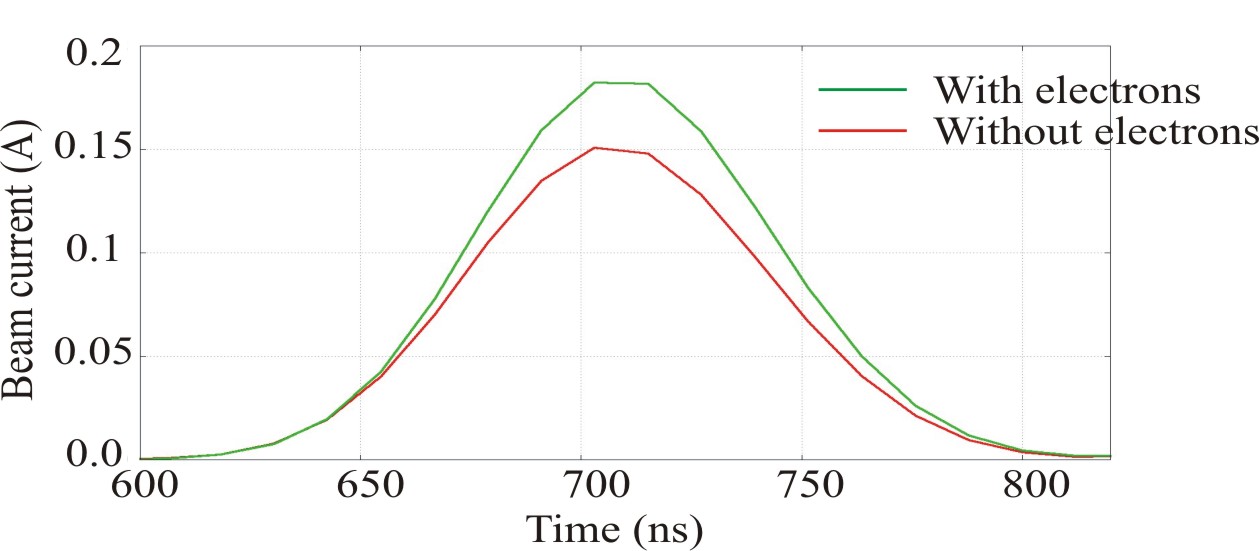}
\end{center}
\caption{Graph showing the space charge compensation by secondary electrons. The beam current is higher when secondary electrons are present reducing beam divergence.}
\label{current_electrons}
\end{figure}
Fig. \ref{current_electrons} show a factor about $1.2$ in the increased beam current due to reduced beam diameter.
The disadvantage may be foreseen as electrons reaching the electric plates causing high voltage break down or sparking.

\subsection{Experiments}
\label{Experiments_chopper}

\begin{figure}[!h]
\vspace{0.5cm}
\begin{center}
\includegraphics*[width=8cm]{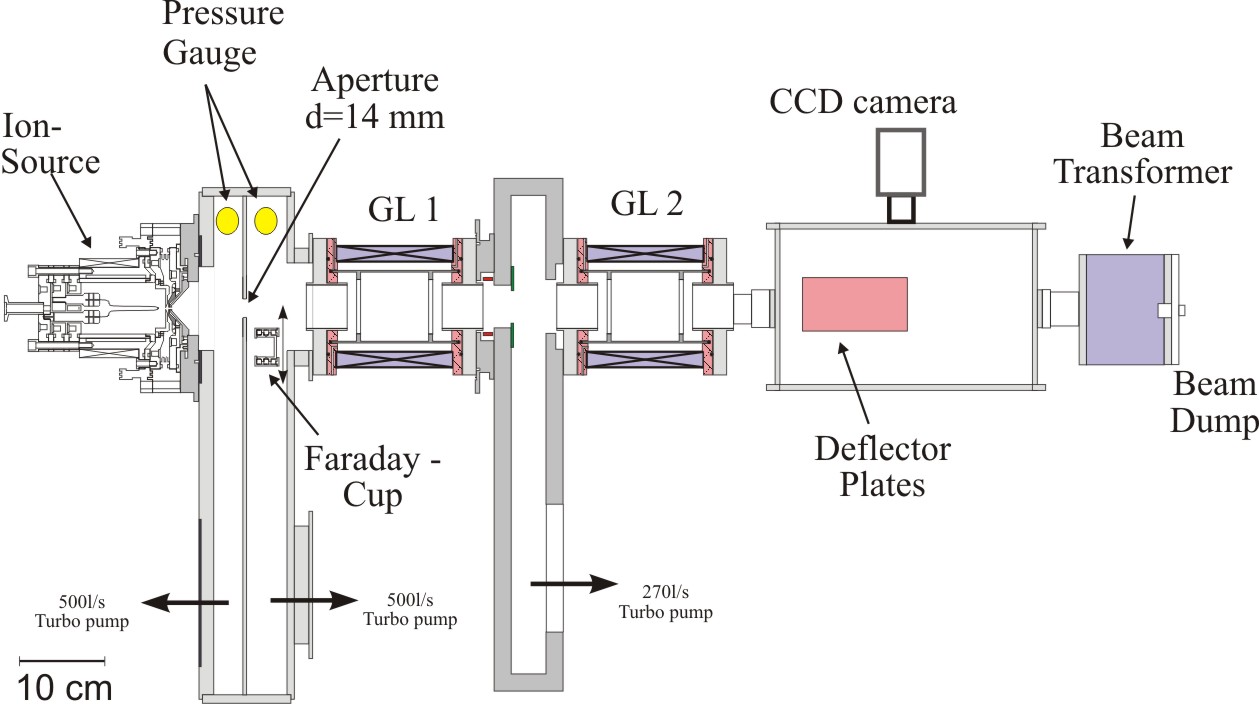}
\end{center}
\caption{Experimental setup for electric kicker experiments.}
\label{kicker_setup}
\end{figure}

The experiments are arranged to test the beam transport through electric kicker and compare the numerical model.
The experimental setup is shown in Fig.~\ref{kicker_setup}.
The $He$-beam (energy $20~keV$, current $1.12~mA$ ) from volume type ion source matched into the deflector tank using an array of two Gabor Lenses (plasma lens GL).
Deflector tank has plates separated by a distance of $38~mm$ over the length of $150~mm$. 
A high voltage pulse generator deflects the beam in the horizontal plane.
Downstream of the deflector tank, a current transformer is installed, followed by the beam dump.

The pulsed electric field deflects the beam away which is recorded as a negative signal in the current transformer.
The current transformer also shows the presence of positive peaks, before and after the beam signal.

\begin{figure}[!h]
\vspace{0.5cm}
\begin{center} 
\includegraphics*[width=8cm]{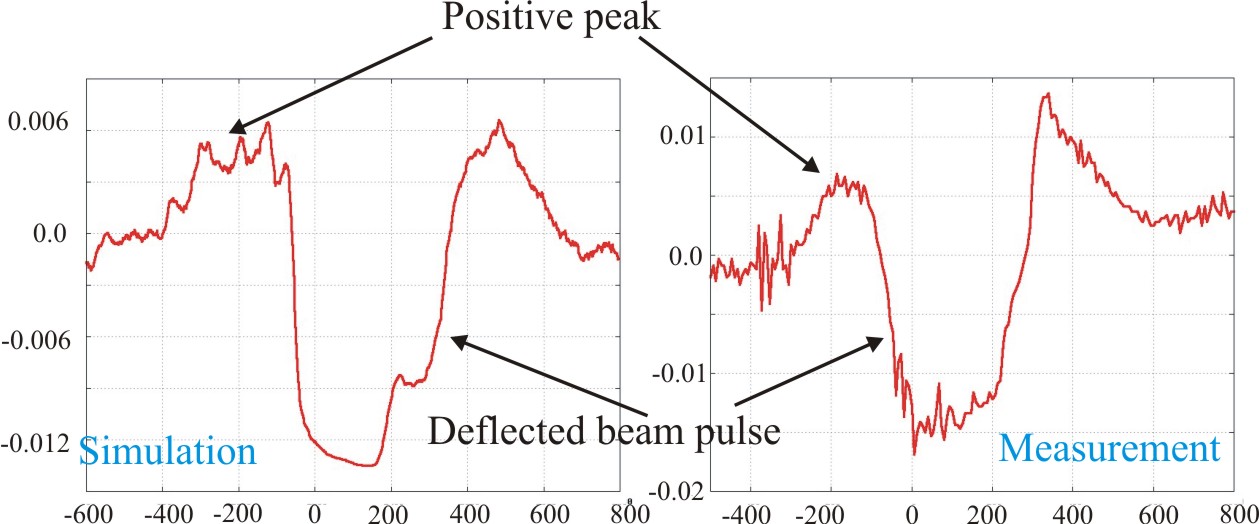}
\end{center}
\caption{Qualitative comparison of simulation with measurement showing beam pulseas a negative signal and the presence of positive peaks indicating the influence of rest gas electrons and secondary electrons.}
\label{kicker_result}
\end{figure}

Int simulation model the compensation electrons (CE) were defined, with homogenous distribution along the beam line.
The secondary electrons are produced at the beam dump.
A single electron was defined to produced in the opposite direction of the beam for every $He$ - ion hitting the beam dump.
This gives an electron flow in the opposite direction causing space charge compensation.
A steady flow of particles gives a zero current in the current transformer.
As the pulse comes in the electrons are flown away before the ions and similarly are again trapped in the beam potential at the end of the pulse.
This gives rise to two positive peaks.
This scenario can be compared in Fig.~\ref{kicker_result}.
The positive peaks due to CE and secondary electrons can be clearly seen although the amplitude does not match exactly.

\section {Monte Carlo Collision (MCC) subroutine}
\label {Monte Carlo Collision (MCC) subroutine}

Particle in Cell model is deterministic where as Monte Carlo method is probabilistic.
The PIC is used to simulate collective effect by calculation of self electric field and external fields and particles are moved through small time step.
In the MC model the probability of collision and collision frequency is calculated using cross sectional probability as a function of energy, target density and time step.

The kinetic energy of a particle, say of the specie $s$ ,
	\begin{equation}
	\mathcal{E}_i = \frac{1}{2} m_s v^2
	\end{equation}
gives the total collision cross section 
	\begin{equation}
	\sigma_T(\mathcal{E}_i) = \sum _{j}\sigma_j(\mathcal{E}_i)
	\end{equation}
where $i$ is the particle number and $j$ is the type of collision.
Then the collision probability for the $i$th particle is given by,
	\begin{equation}
	P_i= 1 - exp( -\bigtriangleup t~v_i~\sigma_T(\mathcal{E}_i)~n_{target}(\mathbf{r}_i))
	\end{equation}
where $\bigtriangleup t$ is the time step and $n_{target}(\mathbf{r}_i)$ is the density of target specie.

Four random numbers are then defined say $R_1,R_2, R_3, R_4$.
First two numbers determines that if the collision occurs or not and which type of collision process takes place.
If $R_1\in[0,1]$ such that $R_1<P_i$, the collision is said to take place.
Second random number determines the type of collision.
There can be multiple numbers of processes taking place with cross section probability overlapping.
Only one reaction can be defined in single time step.
Thus random number $R_2$ is used to determine the type of collision process.
$R_3$ and $R_4$ are used e. g. in the case of electron-neutral collisions

\begin{figure}[!h]
\vspace{0.5cm}
\begin{center}
\includegraphics*[width=8cm]{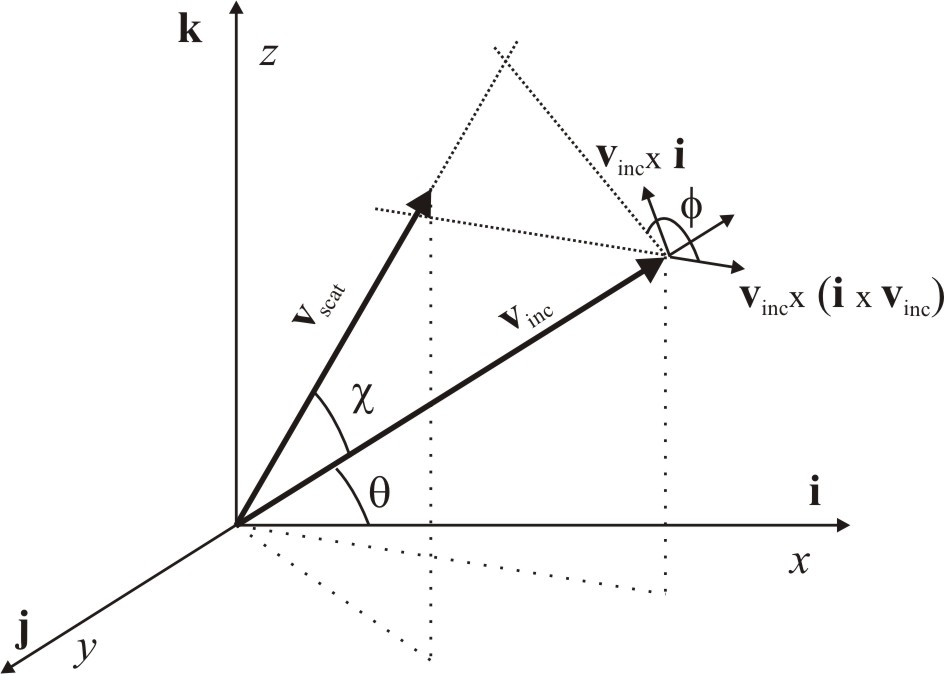}
\end{center}
\caption{Vector diagram showing incident and scattering velocity.}
\label{vectors}
\end{figure}

In electron neutral scattering collision the incident electron is scattered though say $\chi$ (see Fig. \ref{vectors}). 
Then the energy and $\chi$ are related by,
	\begin{equation}
	cos~\chi =\frac{2+\mathcal{E}-2(1-\mathcal{E})^{R_3}}{\mathcal{E}}
	\end{equation}
where $R_3\in[0,1]$ is the random number.
The azimuthal scattering angle $\phi$ is assumed uniform over $[0,1]$ and is given by
	\begin{equation}
	\phi=2 \pi R_4,
	\end{equation}
where $R_4\in[0,1]$ is the last random number.
With similar procedure, the cross sections for different type of reaction can be calculated.

This MC model is parallel implemented in the simulation package to study simple case of He-plasma in volume type ion source.
The initial results are only used to study the correlation of beam emittance and the plasma density in given set of electrode.
In future end to end ion beam transport can be simulated for specific experiments.

\section {Conclusions}
\label {Conclusions}

In this paper we have described features of the simulation package employing a PIC model development.
This numerical model is calculates the beam optics for space charge dominated transport with realistic external fields.
The further development will concentrate on the efficiency of the model and the parallelization for faster calculation.
For the magnetic storage ring project the model is adopted to design the injection system.
The injection system is under construction with a main beam line already installed.
The chopper system with curved electric plates has been designed and installed in the LEBT section of FRANZ beam line.
The Monte Carlo subroutine will be further developed for the time resolved space charge compensation in the kicker system of  the FRANZ.
Ion beam interaction with different rest gas molecule will also be included in calculations.
Independent software will be dedicated using MC method for simulation of discharge plasmas in ion sources.

\section {Acknowledgment}
\label {Acknowledgment} 
I would like to thank C. Wiesner, H. Dinter for collaberation and data exchange from experiments.
I would also like to acknowledge the project team working on the Storage ring project F8SR and the FRANZ.

\end{document}